\documentclass[reprint,aps,pre,twocolumn,groupedaddress]{revtex4-1}
\usepackage{graphicx}
\usepackage{amsmath}

\newcommand{\Alignment}{\mathcal{A}}
\newcommand{\seq}{\vec}
\newcommand{\Nseq}{N_{\mathrm{seq}}}
\newcommand{\Nsamp}{N_\mathrm{samples}}
\newcommand{\gapl}{M}
\newcommand{\lW}{\texttt{W}}
\begin{document}

\title{Score distributions of gapped multiple sequence
alignments down to the low-probability tail}

\author{Pascal Fieth}
\email[]{pascal.fieth@uni-oldenburg.de}
\affiliation{Institut f\"ur Physik, Carl von Ossietzky Universit\"at
Oldenburg, D-26111 Oldenburg, Germany}
\author{Alexander K. Hartmann}
\affiliation{Institut f\"ur Physik, Carl von Ossietzky Universit\"at
Oldenburg, D-26111 Oldenburg, Germany}

\date{\today}

\begin{abstract}
Assessing the significance of alignment scores of optimally aligned DNA or amino acid sequences can be achieved via the knowledge of the score distribution of random sequences.
But this requires obtaining the distribution in the biologically relevant high-scoring region, where the probabilities are exponentially small.
For gapless local alignments of infinitely long sequences this distribution is known analytically to follow a Gumbel distribution.
Distributions for gapped local alignments and global alignments of finite lengths can only be obtained numerically.
To obtain result for the small-probability region, specific statistical mechanics-based rare-event algorithms can be applied.
In previous studies, this was achieved for \emph{pairwise} alignments.
They showed that, contrary to results from previous simple sampling studies, strong deviations from the Gumbel distribution occur in case of finite sequence lengths.
Here we extend the studies to the for practical applications in Molecular Biology much more relevant case of \emph{multiple} sequence alignments with gaps.
    We study the distributions of scores over a large range of the support, reaching probabilities as small as $10^{-160}$, for  
global and local (sum-of-pair scores) multiple alignments.
    We find that even after suitable rescaling, eliminating the sequence-length dependence, the distributions for multiple alignment differ from the pairwise alignment case.
Furthermore, we also show that the previously discussed Gaussian correction to the Gumbel distribution needs to be refined, also for the case of pairwise alignments.
\end{abstract}


\maketitle

\section{Introduction}
One important task in bioinformatics is the analysis of nucleotide or amino acid sequences, e.g.~found in DNA, RNA, or proteins.
A vast amount of sequence data exists and can be found in large data bases like PDB \cite{berman_protein_2000} or UniProt\cite{the_uniprot_consortium_uniprot:_2015}.
A particular important group of methods, widely used for queries to such data bases, is \emph{sequence alignment} \cite{durbin_biological_1998,clote_computational_2000}.
Naturally appearing DNA or amino acid sequences are aligned in a way that is most likely to resemble their actual evolutionary relationship.
Such alignments may contain \emph{gaps}, where in the corresponding evolutionary processes insertion or deletion of genetic material occurred.
The degree of similarity, i.e., relatedness, is quantified by a so-called \emph{(optimum) alignment score}. Different alignment algorithms exist, usually employing fast heuristics, like BLAST \cite{altschul_basic_1990}.

Since the alignment score is just a natural number, one needs to assess the significance of an alignment.
This is achieved typically via calculating the cumulative probability $p(S\geq S_O)$ to find a score $S$ bigger than or equal to the observed score $S_O$ within a suitably defined \emph{null ensemble} of random sequences.
    This approach corresponds to the calculation of $p$-values within standard hypothesis tests.
In most cases, like for standard database queries, the significance analysis is based on sequences where the letters are drawn identically and independently distributed (i.i.d.) using given probabilities.
Therefore it is desirable to find the score distribution of alignment scores for the specified null ensemble.
An analytical solution for the score distribution only exists for the {limiting case of} gapless local alignments with infinite sequence lengths \cite{karlin_methods_1990}, which is not so relevant for biological analysis but 
of academic interest.
According to this theory, the probability function $p(S)$ follows a \emph{Gumbel} (or \emph{extreme-value}) distribution:
\begin{equation}
    p_\mathrm{G}(S) = 
    \lambda\exp\left(-\lambda(S-S_0)-e^{-\lambda(S-S_0)}\right).
    \label{eq:gumbel_distribution}
\end{equation}

For local alignments of sequences of finite length and with allowing for gaps in the alignment the distribution has to be analysed numerically.
Simple sampling studies can easily randomly generate and align, e.g., $10^6$ sequence pairs in computationally feasible time.
This leads to sampling the distribution in the high probability region with lowest statistically reliable probabilities still at $p\approx10^{-5}$.
Early numerical studies using such a simple sampling approach indicated the Gumbel distribution to be a good estimate also for gapped alignments of finitely long sequences \cite{altschul_local_1996}.
However, biological sequences are in most cases very similar to each other.
The relevant alignments therefore generally lie in the high-scoring, extremely low-probability tail of the distribution.
Unfortunately, this tail is not covered by the simple sampling approach.
To obtain results for the low-probability region, one of us (AKH) applied a statistical mechanics-based rare-event simulation to the problem of pairwise local sequence alignment \cite{hartmann_sampling_2002}.
This work showed that the Gumbel distribution significantly deviates from the obtained score distribution in the distribution tails.
Subsequent studies indicated that a Gaussian correction to this distribution suits well to gain a better description of the score distribution.
Interestingly, the strength of this corrections decreases with increasing sequence length \cite{wolfsheimer_local_2007}.
Below we will show that this Gumbel distribution with correction factor leads to a better fitting distribution, but is still only an incomplete description of the data.
These methods only have been used so far for pairwise local sequence alignment with and without gaps.
It is the main purpose of this work, to extend the application of these methods to multiple local sequence alignment, where more than two sequences are compared within one alignment.
Multiple sequence alignments are vastly used for analysing three or more sequences with an assumed evolutionary relation.
    Resulting alignments are amongst other things used to estimate the phylogenetic history of sequences or to find highly conserved protein domains.
    Similar protein functions can result from actual evolutionary relation or from convergent evolution, i.e.~similar functions developed in independent branches of the phylogenetic tree.
Significance analyses could help distinguishing the two cases.
Due to the expensiveness of local multiple sequence alignment, its counterpart global multiple sequence alignment is much more common.
However, local multiple sequence alignment is especially suited for finding functionally important regions within whole protein families.

Note that for global sequence alignment no analytical solution for the score distribution exists.
Studies of real datasets \cite{webber_estimation_2001} and subsequent studies following essentially a simple sampling approach \cite{pang_statistical_2005} suggest the three-parameter Gamma distribution as a good model:
\begin{equation}
    p_{\mathrm{gamma}}(S)=
    \begin{cases}
        \frac{\lambda^{\gamma}(S-\mu)^{\gamma-1} e^{-\lambda(S-\mu)}}
        {\Gamma(\gamma)}
        & S>\mu\\
      0 & S\leq \mu,
    \end{cases}
    \label{eq:gammaDistribution}
\end{equation}
with the Gamma function $\Gamma(x)$, and parameters $\lambda$, $\gamma$, and a shift $\mu$.
Due to the nature of the studies, the sampled region was again restricted to the high-probability region of the score distribution.
As a second application of the rare-event simulation, we studied the score distributions of i.i.d.~random sequences in the low-probability region of pairwise and multiple global sequence alignments.

The remainder of this work is organized as follows.
First, we will formally define the alignments we studied and state the alignment algorithms we used.
Next, we explain the statistical mechanics-based large-deviation approach, which allowed us to sample the alignments distribution of random sequences over a large range of the support.
In the main section, we show our results, for gapped multiple alignments of two, three and, in case of global alignment, four  sequences.
The main results are that again the Gumbel distribution is not sufficient to model the data and that the distribution for multiple alignments, even more relevant for practical applications in Molecular Biology, cannot be obtained from the pairwise alignments results, justifying the present numerically demanding study.
Finally, we present a summary and an outlook.

\section{Sequence Alignments \label{sec:alignment}}

Sequence alignment algorithms aim to find the optimal scoring alignment of two or more sequences.
DNA and amino acid sequences are given by a representing sequence of letters from an alphabet $\Sigma$.
It is $\left|\Sigma\right|=4$ for DNA sequences consisting only of the four bases.
In contrast, there are 20 different amino acids, leading to $\left|\Sigma\right|=20$ for the respective sequences.
Let $\seq{x^{(i)}}=x^{(i)}_1,x^{(i)}_2,\dots,x^{(i)}_{L_i} \in \Sigma^{L_i}$ be the $i^\mathrm{th}$ sequence of length $L_i$ in a set of $\Nseq$ sequences.
A multiple alignment $\Alignment$ is defined as a set of tuples of indices
\begin{equation}
    \begin{aligned}
        \Alignment=\{(l^{(1)}_k,l^{(2)}_k,\dots,l^{(\Nseq)}_k)\};\;
        & k=1,2,\dots,K\\
        & 1 \leq l^{(i)}_k < l^{(i)}_{k+1}\leq L_i.
    \end{aligned}
    \label{eq:alignmentDefinition}
\end{equation}
A pair of letters $(x^{(i)}_{l^{(i)}_k},x^{(j)}_{l^{(j)}_k})$ is called a \emph{match} if $x^{(i)}_{l^{(i)}_k}=x^{(j)}_{l^{(j)}_k}$.
Otherwise it is called a \emph{mismatch}.
Note that in each tuple $\Nseq$ letters are joined, so some pairs in this tuple may match while other pairs may form a mismatch.
If $l^{(i)}_{k+1}=l^{(i)}_k+1$ and $l^{(j)}_{k+1}=l^{(j)}_k+1+\gapl$ with $\gapl>0$, i.e., the indices $l^{(j)}_k+1,\ldots, l^{(j)}_k+M$ do not appear in any tuple, sequence $\seq{x^{(i)}}$ is said to have a \emph{gap} of length $\gapl$ in respect to
sequence $\seq{x^{(j)}}$. Also if $l^{(k)}_1=1+M>1$ or $l^{(k)}_K=L_i-M<L_i$ one speaks of a gap of the $k$'th sequence.

To find the alignment most likely resembling the actual evolutionary relationship, an objective \emph{score} function $S(\Alignment,\{\seq{x^{(i)}}\})$ is used.
Usually, the score is based on sums of pairwise scores $s(x^{(i)}_{l^{(i)}_k},x^{(j)}_{l^{(j)}_k})$, which are taken from so-called \emph{substitution matrices} given by biologists.
Typically, the scores are proportional to $\log$ of mutation (mismatch) or conservation (match) probabilities in evolutionary models.
Therefore, scores for matches are positive, while scores for mismatches are smaller, usually negative. 
Two widely used sets of substitution matrices are the PAM \cite{dayhoff_chapter_1978} and BLOSUM \cite{henikoff_amino_1992} substitution matrix sets.
Gaps are penalised depending on their lengths with some length-dependent function $g(\gapl)$.
The gap penalties are intended to model log of probabilities of the occurrence of insertion or deletion events in evolutionary processes.

To wrap this up, the score of a \emph{global} pairwise alignment is the sum over all matches, mismatches, and the penalties of all gaps $g$.
A multiple alignment is scored by the sum-of-pairs score, i.e., its score is the sum over all pairwise scores:
\begin{equation}
    S(\Alignment,\{\seq{x^{(i)}}\})=\sum_{i<j}\left(\sum_k
    s(x^{(i)}_{l^{(i)}_k},x^{(j)}_{l^{(j)}_k})-\sum_g
g(\gapl_g)\right)\,,
    \label{eq:alignmentScore}
\end{equation} 
where the sum $\sum_g$ runs over all gaps.

The aim of the alignment comparison is that the more similar two sequences are, the higher the resulting score.
Nevertheless, the score does not depend only on the sequences but also on the alignment.
For example, even for two equal sequences, one can find alignments with a very small score.
Thus, one is seeking for the \emph{optimal alignment} $\Alignment_O$ which is the alignment with the maximum score

$$S_\mathrm{max}(\{\seq{x^{(i)}}\})=
\max_{\Alignment}S(\Alignment,\{\seq{x^{(i)}}\})\,,$$ 
maximized over all possible alignments.
In subsequent sections $S$ is used synonymously with $S_\mathrm{max}(\{\seq{x^{(i)}}\})$.
In this work, affine gap costs 
\begin{equation}
g(\gapl)=\alpha+\beta(\gapl-1)
\label{eq:affine}
\end{equation}
were used.
Thus, a high penalty $\alpha$ can be given for opening a gap and a smaller penalty $\beta$ for extending one.
Although even for two sequences there are exponentially many alignments, an optimal pairwise global alignment, i.e., $\Nseq=2$, with affine gap costs can easily be found in polynomial time by the algorithm by Needleman and Wunsch\cite{needleman_general_1970} with Gotoh's extension\cite{gotoh_improved_1982}.
For multiple global alignments the \emph{progressive} heuristics by Feng and Doolittle \cite{feng_progressive_1987} was used in this work:
All possible sequence pairs are aligned with pairwise alignment first.
A guide tree is constructed using the different obtained pairwise scores.
Sub-alignments are then aligned to each other by aligning the highest-scoring sequence pair, taking into account the existing alignments.

Another important alignment method next to global sequence alignment is \emph{local} sequence alignment.
Here, from each sequence a subsequence is chosen and only the subsequences are aligned.
This means for each sequence start- and endpoints $l^{(i)}_s$, $l^{(i)}_e$ of the subsequences are subject to optimization as well.
Thus, the optimal local alignment has to maximise the score over all possible alignments of all possible subsequences.
This is also equivalent to not penalising gaps at the beginning and the end of sequences.
With the algorithm by Smith and Waterman\cite{smith_identification_1981} this is possible also in polynomial time $\mathcal{O}(L^{\Nseq})$ for a set of $\Nseq$ sequences, if each has the length $L_i=L$.

\section{Method\label{sec:method}}
\begin{figure}
    \includegraphics[width=0.9\columnwidth]{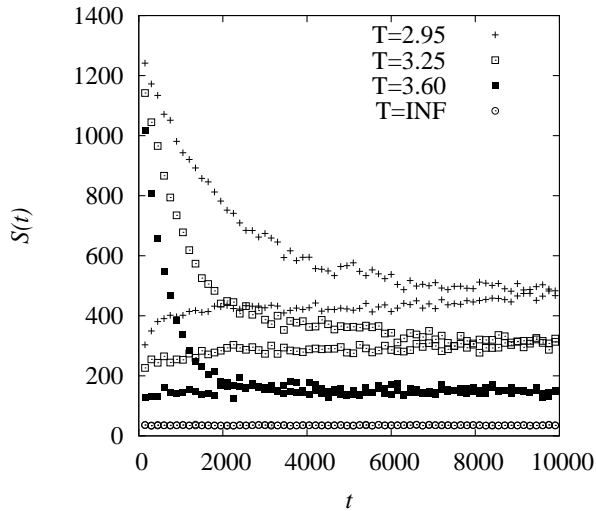}
    \caption{Equilibration times for different temperatures:
        First 10,000 Monte Carlo sweeps in the parallel tempering 
 time series averaged over 30 different realisations for random 
(bottom branch) and 
high scoring (top branch) initial conditions, respectively.
        System:
        $\Nseq=3$ sequences of length $L=40$, local alignment.
        A selection of the used temperatures is shown.
At lower temperatures it takes more time for equilibration.
    \label{fig:equilibration}}
\end{figure}

In the approach used \cite{hartmann_sampling_2002}, a sequence pair, correspondingly here a set of $K$ sequences, is viewed as a ``state'' $\mathcal{C}$ of a physical system with ``energy'' $E=-S$.
A ``temperature'' $T$ is introduced and the states are sampled according to the rules of the canonical ensemble in statistical mechanics.
Specifically, a Markov chain $\mathcal{C}_0 \rightarrow \mathcal{C}_1 \rightarrow \dots$ is generated.
In each step $t$ a trial state $\mathcal{C}'$ is generated from the current state $\mathcal{C}_t$ by randomly choosing and replacing one residue in the sequence set \footnote{ For amino acid sequences the new residue is drawn according to the naturally occurring background frequency as found by Robinson \& Robinson \cite{robinson_distribution_1991} }.
The alignment score $S(\mathcal{C}')$ is calculated and the trial state accepted, i.e., $\mathcal{C}_{t+1}=\mathcal{C}'$, with the Metropolis probability \cite{metropolis_equation_1953} $P(\mathcal{C}_t\rightarrow\mathcal{C}')=\min\left[1,\exp(\Delta S/T)\right]$ with $\Delta S=S(\mathcal{C}')-S(\mathcal{C}_t)$.
If not accepted, the current configuration is kept, i.e., $\mathcal{C}_{t+1}=\mathcal{C}_t$.
The equilibrium distribution of the sampled states is known to be $Q(\mathcal{C}) = P(\mathcal{C})\exp\left({S(C)/T}\right)/Z_{T}$ with the partition function $Z_{T} = \sum_\mathcal{C}P(\mathcal{C})\exp\left(S(\mathcal{C})/T\right)$.
The score distribution, biased by the scale parameter (or ``temperature'') $T$ is then:
\begin{eqnarray*} p_T(S) & = &
    \sum_{\{C|S(C)=S\}} Q(C)\\
     & = & 
    \sum_{\{C|S(C)=S\}} \frac{\exp\left(S/T\right)}{Z_T}P(C)\\
    & = & \frac{\exp\left(S/T\right)}{Z_T}p(S)
    %\label{eq:biased_distribution}
\end{eqnarray*}
The unbiased distribution then is $p(S) = p_T(S)Z_T\exp(-S/T)$.
After rescaling via $p^R_T=p_T\exp(-S/T)$ only the correct values for the partition functions $Z_T$ remain unknown.
Determining them is possible after covering different probability regions, starting with the one for $T=\infty$.
To cover the entire score range, simulations are done at different temperatures.
The efficiency of simulations is then improved further by using the parallel tempering technique \cite{geyer1991,marinari_simulated_1992,hukushima_exchange_1996}, which works as follows.
Simulations of the system are done for $N_T$ different temperatures $T_1<T_2<\dots<T_{N_T}$, i.e., an independent configuration $\mathcal{C}_{i}$ is simulated for each temperature $T_i$, $i=1,\ldots,N_T$.
The configurations at neighbouring temperatures $(T_i,T_{i+1})$ are swapped in suitably chosen simulation time intervals with a swap probability $P_\mathrm{sw} (\mathcal{C}_i\leftrightarrow\mathcal{C}_{i+1}) = \min(1,\exp[-\Delta \beta\Delta S])$ with $\Delta \beta = 1/T_{i+1}-1/T_i$ and $\Delta S = S(\mathcal{C}_{i+1})-S(\mathcal{C}_i)$.
One time step $t$ consists here of one Monte Carlo sweep, i.e., a number of Metropolis steps equal to the total number of residues in a configuration, and a subsequent sweep of $N_T-1$ exchange attempts of randomly selected temperature pairs.
To sample only weakly correlated data, the autocorrelation $c_S(t) = \{\left<S(t_0)S(t_0+t)\right> - \left<S\right>^2\} / \{\left<S^2\right> - \left<S\right>^2\}$ is calculated and a relaxation time $\tau$ determined for which $c_S(\tau)=1/e$.
Only every $\tau^\mathrm{th}$ value is considered when sampling the data.
Equilibrium is ensured by starting to sample after an equilibration time $t_0$.
To determine this time, two sets of initial conditions are simulated.
One set of simulations is initialised with randomly generated sequence sets, i.e., with low scoring sequence sets.
The other set of simulations is initialised with sets of identical sequences, only consisting of the letter $a$ with the highest pairwise score $s(a,a)$ \footnote{For the BLOSUM62 matrix this is the amino acid tryptophan (\lW) with $s(\lW,\lW)=11$ }, i.e., with maximum scoring sequence sets.
The equilibration time $t_0$ is reached when the average scores of the different regions have reached the same value within error bars.
Compare Fig.~\ref{fig:equilibration} which shows the first 10,000 time steps of the simulations for the simulations done for $\Nseq=3$ sequences of lengths $L=40$ for different temperatures $T$.

Having covered the whole distribution range, it is now possible to determine the partition functions $Z_T$.
Assuming for $T=\infty$ the biased distribution approximates the true distribution $P_T(S)\approx P(S)$ the partition function, or normalisation constant, becomes $Z_\infty \approx 1$.
After rescaling, the other distributions differ from the unbiased distributions only by $Z_T$ which can be determined by a shift between neighboring 
functions $T_{i-1,i}$ on the logarithmic scale, $\log Z_i$.
The shift between $T_1=\infty$ and the unbiased distribution is defined as 
$\log Z_1=0$.
The consecutive shifts are determined by minimising the function
\begin{equation}
    \begin{split}
        f(\log Z_i)=\sum_{S\in[S_{s},S_{e}]}\{&\left(\log
        p_{T_i}^R(S)+\log Z_i\right)-\\
        &\left(\log p_{T_{i-1}}^R(S) + \log Z_{i-1}\right) 
        \}^2.
    \end{split}
    \label{eq:findShift_min}
\end{equation}
in the overlapping score region $[S_s,S_e]$ of the two neighboring distributions.
The shifted distributions are then $p^S_T = p^R_T\cdot Z_T$.
For most score ranges, several values from the different biased distributions are available.
For each value $S$ of the score distribution a weighted average over all available data is calculated:
\begin{equation}
    p(S)=\frac{1}{\sum_{T}w_T(S)}\sum_{T}w_T(S)p^S_T.
    \label{eq:distribution_weightedAvg}
\end{equation}
$w_T$ is the inverse relative error of the corresponding distribution $p_T(S)$ before rescaling and shifting:
\begin{equation}
    w_T(S)=\sqrt{\frac{n_T}{p_T(S)(1-p_T(S))}}
    \label{eq:mergeWeights}
\end{equation}
where $n_T$ is the number of used samples for temperature $T$.
This finally gives values for a wide range of the distribution $p(S)$ down to regions of very low probabilities.

\section{Results\label{sec:results}}
The rare-event simulations were performed for local and global alignments.
If not mentioned otherwise, the alignments were calculated using the BLOSUM62 substitution matrices and gap penalties according to \eqref{eq:affine} with $(\alpha=12, \beta=1)$.
These values were used in the previous large-deviation studies \cite{hartmann_sampling_2002, wolfsheimer_local_2007} and therefore allow a direct comparison between results for pairwise and multiple sequence alignment.

\subsection{Local multiple sequence alignments}
Rare-event simulations for local pairwise sequence alignments have shown that the Gumbel distribution alone is not a good description of the score distribution of gapped local sequence alignments.
The introduction of a Gaussian correction to the Gumbel distribution \eqref{eq:gumbel_distribution} improved the agreement between analytical distribution and data.
The corrected distribution is
\begin{equation}
    \begin{aligned}
        p_\mathrm{C} & =  p_\mathrm{G}\cdot\exp[-\lambda_2(S-S_0)^2]\\
                     & = \lambda 
        \exp\left(-\lambda(S-S_0) - \lambda_2(S-S_0)^2 -
        e^{-\lambda(S-S_0)}\right).
    \end{aligned}
    \label{eq:gumbel_gauss_distribution}
\end{equation}
The strength of the Gaussian correction is indicated by the fit parameter $\lambda_2$ and has been shown to decrease with increasing sequence length for pairwise alignments.
$\lambda_2$ was observed to decrease with a power-law for small gap costs and faster than a power-law for high gap costs.
In the case analysed here, ($\alpha=12$, $\beta=1$), the decrease is expected to be just in the power-law region.
Subsequently the distributions obtained with the rare-event simulations for multiple local sequence alignment of $\Nseq=3$ sequences will be presented and later on compared to the results for pairwise alignment.

\subsubsection{Score distributions for local multiple sequence alignments}

\begin{figure}
    \includegraphics[width=0.9\columnwidth]{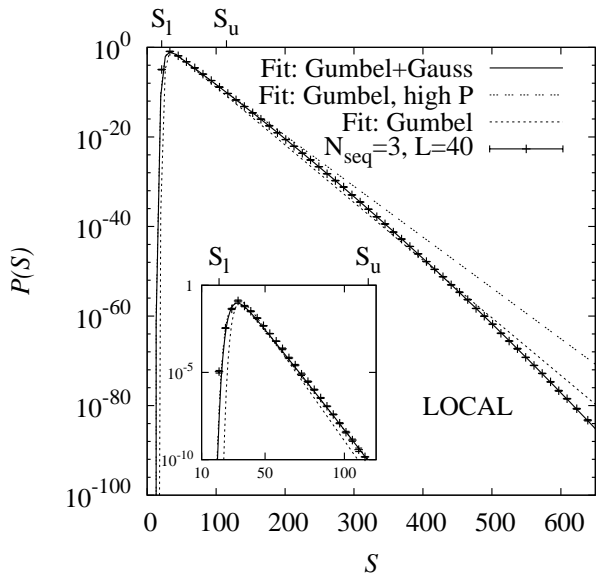}%
    \caption{
        The distribution for local alignments of $\Nseq=3$ sequences of 
length $L=40$ as obtained by the rare-event simulation.
        Fits of the Gumbel distribution to the whole distribution as 
well as constrained to the score range 
$[S_l=21,S_u = 115]$, i.e., $P(S)\leq 10^{-10}$, are shown.
        Also shown is the better suited fit of the Gumbel distribution with a Gaussian correction.
        The inset shows the high probability region.
        \label{fig:localDistributionCompare}}
\end{figure}

Simulations for local multiple sequence alignments of $\Nseq=3$ sequences of length $L=40$ were performed for 60 different realisations of the driving randomness, each over $5\cdot10^4$ sweeps.
The score distribution was obtained as described in sec. \ref{sec:method}.
A fit of the Gumbel distribution without \eqref{eq:gumbel_distribution} and with \eqref{eq:gumbel_gauss_distribution} Gaussian correction was performed.
The data and the fits are shown in Fig.~\ref{fig:localDistributionCompare}.
The deviation of the Gumbel distribution from the data in the tail of the distribution is clearly visible.
The better performance of the fit of the Gumbel distribution with Gaussian correction is also indicated by the $\chi^2$ value per degree of freedom, $\chi^2/\mathrm{ndf}=1.5$ in contrast to the value for the distribution without Gaussian correction, $\chi^2/\mathrm{ndf}=343.1$.
Thus, the behavior for the low-probability tail for pairwise alignment could be confirmed for the alignment of $\Nseq=3$ sequences.
We performed another fit where  the Gumbel distribution without Gaussian correction was restricted to the distribution range with $P(S) \geq 10^{-10}$, which corresponds to the interval $[S_l=21, S_u=115]$.
Simple sampling would only cover probabilities $P(S) > \Nsamp^{-1}$ by creating and aligning $\Nsamp$ i.i.d.~random sequence sets.
Therefore, the restricted distribution range is still generously large, requiring more than $\Nsamp>10^{10}$ samples if it were to be obtained by simple sampling.
The restricted fit agrees with the data in the high-probability region, visible in the inset.
This is also indicated by a somehow better $\chi^2$-value of $\chi^2/\mathrm{ndf}=63$.
Of course the fit now obviously deviates strongly in the distribution tail.
Indeed, the Gumbel distribution with correction constrained to the same score range still performs better than without correction ($\chi^2/\mathrm{ndf}=9.45$).
Over the complete obtained distribution range this results in a significant overestimation of $\lambda_2$ and a strong underestimation of $P(S)$ in the tail of the distribution.
The results of these fits show that the behaviour of the distribution can not be fully studied with just a simple sampling approach.

\begin{figure}
    \includegraphics[width=0.9\columnwidth]{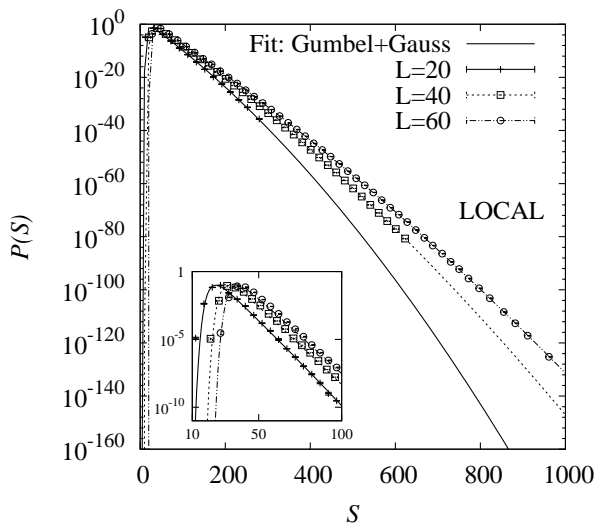}%
\caption{
    The score distributions for local multiple alignments for several sequence lengths and the fitted Gumbel distributions with Gaussian correction.
    The inset shows the high-probability region.
    \label{fig:distributionsGathered}}
\end{figure}

\begin{figure}
    \includegraphics[width=0.9\columnwidth]{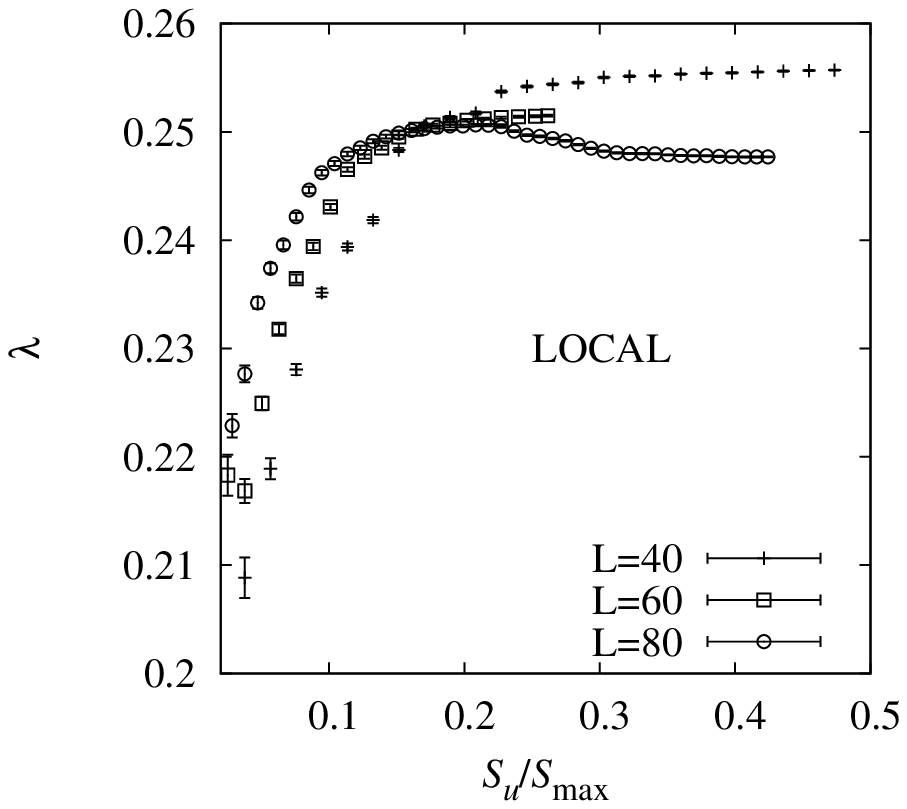}%
\caption{The fit parameters found for different window sizes $(S_\text{min},S_\text{end})$.
Several sequence lengths $L$ for $\Nseq=3$ are shown.
    \label{fig:lambdaWindow}}
\end{figure}

Figure \ref{fig:distributionsGathered} shows the fit of the Gumbel distribution with Gaussian correction to the data for local alignment of $\Nseq=3$ sequences for different sequence lengths.
The fit performed as well as for the example case of $L=40$.
However, varying the range of the distribution on which the fit is performed results in a change of parameter values not accounted for by the standard error calculated during the fit.
Therefore parameter fits for every system were performed with varying window sizes.
All windows start at the minimum score $S_l$ for which the simulations yielded a data point (i.e., this point is fixed for each system) and ends at a variablescore $S_u\leq S_{\max}$, where $S_{\max}$ is the maximum possible score, 
i.e., for two equal sequences with the highest scoring letter. 
For the analysed system sizes for multiple local sequence alignment with $\Nseq=3$ the parameters seemed to converge.
Figure \ref{fig:lambdaWindow} exemplarily shows the parameter curve for $\lambda(S_u)$.
For sequence lengths up to $L=60$ it seems reasonable to estimate the parameters by fitting an exponential function with a constant.
For a parameter $g$ we use $g(S_u)=g_b+C\exp(-k\cdot S_u)$ with fit parameters $g_b$, $C$, and $k$ for an appropriate part of the value curve obtained.
The new fit parameter $g_b$ approximates the value against which parameter $g(S_u)$ converges.
For $L=80$ this approach already appears less promising. Restricting the 
window for the fit is necessary to gain a feasible value for the parameters.
Parameter values were obtained in the way described and are shown in 
Tab.~\ref{tab:local_fitparams}, including the range of windows chosen, 
but should be taken with the according caution.
We will see that this lack of confidence is not the result of expanding the analyses to multiple sequence alignment.
The same analysis of the distributions found for pairwise alignments rather shows that the fit performs even worse there.
The analytical solution \cite{karlin_methods_1990} does assume infinitely long sequences.
Generally a decrease of the parameter $\lambda_2$, indicating the strength of the Gaussian correction, can be observed, see Tab.\ \ref{tab:local_fitparams}.
The decrease of $\lambda_2$ with increasing sequence length indicates that the correction is partially due to a finite size effect, disappearing for long sequences.
This decrease has already been found in the study of pairwise alignment distributions \cite{wolfsheimer_local_2007} and can be confirmed here for multiple alignments.
As even the corrected function and the acquisition of its parameters is still quite makeshift, as visible from the $\chi^2/$nfd values shown in tab.\ \ref{tab:local_fitparams}, beginning considerably larger than one, we did not conduct a more quantitative analysis of the decrease of $\lambda_2$.

\begin{table}
    \centering
    \begin{tabular}{c c c c c c}
        \hline\hline
      L & $\lambda$ & $10^{4}\lambda_2$ & $S_0$ &
        $\frac{\chi^2}{\text{ndf}}$ & $S_l$ $[S_l/S_{\max}]$\\\hline
     40 & $0.25570(7)$
        & $0.968(4)$
        & $32.163(4)$
        & 8.2
        & 200$[0.152]$\\

     50 & $0.25378(5)$
        & $0.7441(15)$
        & $35.568(4)$
        & 7.3
        & 225$[0.14]$\\

     60 & $0.25164(6)$
        & $0.6045(17)$
        & $37.9981(13)$
        & 6.7
        & 225$[0.114]$\\

     80 & $0.247685(25)$
        & $0.4889(4)$
        & $41.744(7)$
        & 3.7
        & 750$[0.284]$\\
        \hline\hline

    \end{tabular}
    \caption{Parameters for the fit of the Gumbel distribution with Gaussian
        correction to the obtained data for local multiple sequence alignments\ ($\Nseq=3$).
        Parameter values are convergence values: A fit was performed for each parameter value as a function of the window size of the distribution.
    $\chi^2$ values were calculated for these convergence values of the parameters in the Gumbel distribution in respect to the obtained distribution data.
\label{tab:local_fitparams}}
\end{table}

\subsubsection{Comparison to pairwise alignment}

\begin{figure}
    \includegraphics[width=0.9\columnwidth]{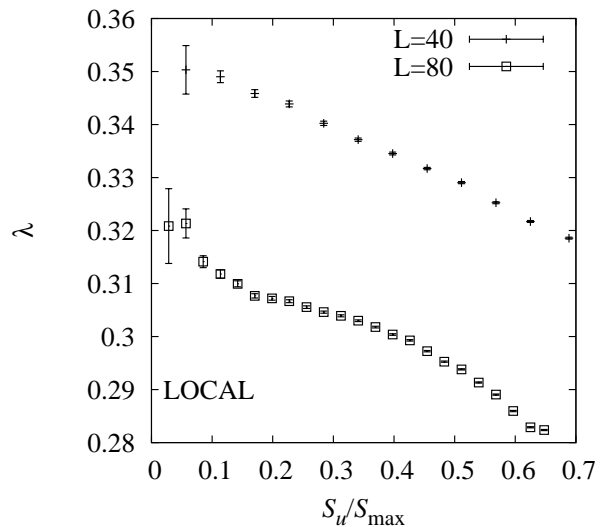}%
\caption{The fit parameters found for different window sizes 
$(S_\text{min},S_\text{end})$. Several sequence lengths $L$ for 
$\Nseq=2$ are shown.
    \label{fig:lambdaWindowPW}}
\end{figure}

As the simulations for multiple sequence alignment were computationally expensive especially in the case of local alignments, only short sequences up to $L=80$ could be used for score distribution analysis in the present work.
For a better comparison with the data for pairwise alignment, where mostly longer sequences were studied \cite{wolfsheimer_local_2007} we have performed additional simulations for shorter sequences also for pairwise alignments.
As we implemented a parallel version of the algorithm for the MSA simulations, we could also obtain the distribution of scores over a larger range of the support, compared to the previous work.
We fitted the data to the Gaussian-corrected Gumbel distribution.
Obtained values are shown in Tab.~\ref{tab:local_fitparams}.
Restricting the fit to the range of the support, which was 
addressed in the previous work where known, reproduces the parameter values found before.
But when we extended the distribution range for the fit further, 
it yielded different parameter values.
Figure \ref{fig:lambdaWindowPW} shows the fit value of $\lambda$ for $\Nseq=2$ for different window sizes.
This resulted in a small but visible linear decrease of the parameter value, showing that in fact even the corrected Gumbel distribution is not sufficient to describe the obtained data over a large range of the support.
The lower boundary of the window, $S_l$, was varied as well, eliminating small $S$ values.
The linear decrease of parameters could be observed for varying $S_l$ while fixing $S_u=S_{\max}$ as well as for varying $S_u$ again, but with different, higher, boundaries $S_l$.
Furthermore, $\chi^2$ values were obtained for actually performing the fit to the whole support obtained and are shown together with parameter values in Tab. \ref{tab:local_fitparamsPW}.
They also indicate that the fit of the corrected Gumbel distribution does not perform as well for pairwise local sequence alignment as for multiple local sequence alignment.

\begin{table}
    \centering
    \begin{tabular}{c c c c c c}
        \hline\hline
      L & $\lambda$ & $10^{4}\lambda_2$ & $S_0$ &
        $\frac{\chi^2}{\text{ndf}}$ \\\hline
     30 & $0.3243(4)$
        & $12.2506(13)$
        & $12.755(29)$
        & 21.2\\

     40 & $0.31857(18)$
        & $9.175(7)$
        & $15.096(19)$
        & 26.5\\

     50 & $0.30084(2)$
        & $7.597(5)$
        & $15.50(3)$
        & 34.5\\

     60 & $0.29452(12)$
        & $6.3636(27)$
        & $16.675(23)$
        & 40.9\\

     80 & $0.2824(12)$
        & $4.8321(17)$
        & $17.42(4)$
        & 43.4\\
        \hline\hline

    \end{tabular}
    \caption{Parameters for the fit of the Gumbel distribution with Gaussian
        correction to the obtained data for local pairwise sequence alignments ($\Nseq=2$).
    As no convergence of parameter values were observed, all values, including $\chi^2$, are given as obtained by fitting over the whole range of the support for which data has been obtained.
\label{tab:local_fitparamsPW}}
\end{table}

\begin{figure}
    \includegraphics[width=0.9\columnwidth]{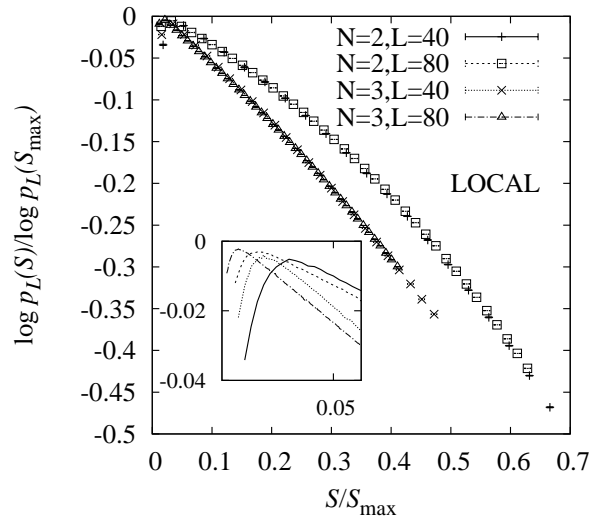}%
\caption{The distributions for various $L$ and $\Nseq$ with rescaled scores and probabilities.
The distributions coincide for low probabilities and identical number of sequences $\Nseq$ but differ among pairwise and multiple alignments.
Inset: High probability region in which the distributions deviate from each other (even for same $\Nseq$).
    \label{fig:localCompareRescaled}}
\end{figure}

Thus, instead of comparing parameters of fitted distributions, 
we rather compare the data itself.
This requires some scaling to get rid of sequence-length dependencies.
The results in Ref.\ \cite{newberg_significance_2008} indicate that for lengths $(L,L')$ of a sequence pair, by rescaling the $S$-axis with $S_{\max}$ and by rescaling the log probabilities by $\log p(S_\text{max}|L,L')$ the curves for different pairs of sequence lengths fall approximately on top of each other.
This means
\begin{equation}
\frac{\log p(S|L,L')}{\log p(S_\text{max}|L,L')} \approx f(S/S_\text{max})\, 
\end{equation}
where $f(.)$ is a universal function of the relative score $Q=S/S_\text{max}$.
Thus, for sequence lengths $\hat{L},\hat{L'}$ and a different score $\hat{S}$ but with the same relative score $Q$, i.e., $\hat{S}=S/S_\text{max}\hat{S}_\text{max}$ one obtains 
\begin{equation}
    \log p(S|L,L') \approx \log p\left(
\frac{S \hat{S}_\text{max}}{S_\text{max}}\Big|\hat{L},\hat{L'}\right)
\frac{\log p(S_\text{max}|L,L')}{\log p(\hat{S}_\text{max}|\hat{L},\hat{L'})}.
    \label{eq:significanceRelation}
\end{equation} 
Figure \ref{fig:localCompareRescaled} shows selected distributions for $\Nseq=2$ and $\Nseq=3$ with rescaled score- and probability axes.
The distributions for pairwise alignments coincide in the low probability region as well as the distributions for multiple alignments.
There is only a deviation in the small region of high-probabilities which disappears with increasing sequences length.
There is however an overall difference between pairwise and multiple alignments.
The distributions do not coincide and the curvature is significantly stronger for $\Nseq=3$.
Thus, the deviation from the Gumbel distributions is stronger for multiple alignments and distributions of scores for multiple alignments can not be estimated easily from the data obtained for pairwise alignment distributions.
Nevertheless, the  knowledge of one distribution of arbitrary length $L$ makes it possible to estimate distributions for other sequence lengths $L'$
also for multiple alignment of $\Nseq=3$ sequences.
The same can be expected for $\Nseq>3$.
But a more precise analysis would require more simulation work for multiple sequence alignments with more sequences.
These analyses could potentially yield a method to estimate the distributions for $\Nseq>3$ by readily obtained distributions with less sequences.
Nevertheless, for an actual study the numerical demand for the alignment of even more sequences appears to be too high.
Here, the implementation of a good working heuristics for multiple local sequence alignments with gaps would be helpful.

\subsection{Global multiple sequence alignments}
We could adopt the large-deviation approach easily to global pairwise and multiple sequence alignment with and without gaps.
We below first present the analysis of the high-probability tail for pairwise sequence alignment and compare it to previous simple sampling results.
Furthermore, we expanded the method to multiple sequence alignment and compare to the results for pairwise sequence alignment.

\subsubsection{Pairwise global sequence alignments} 
\begin{figure}
    \includegraphics[width=0.9\columnwidth]{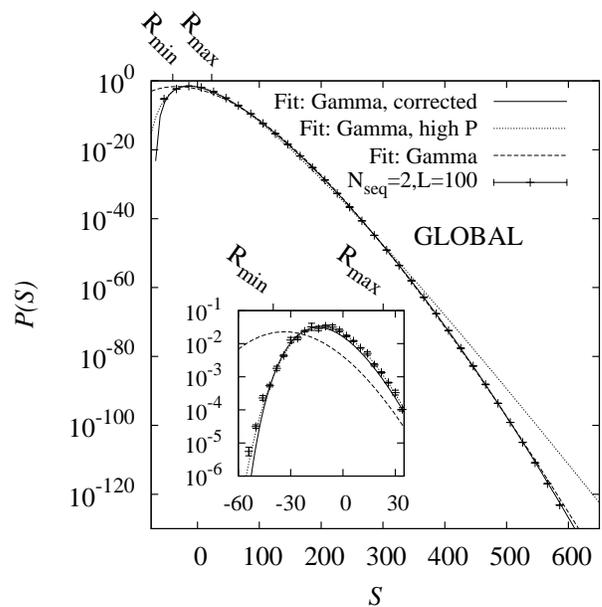}%
    \caption{The alignment score distribution as obtained with the 
rare-event simulations for the global alignment of $\Nseq=2$ 
sequences of length $L=100$, here with gap costs $(\alpha=7,\beta=-1)$ 
for direct comparison with \cite{pang_statistical_2005}.
        Shown are the fits of the Gamma distribution with and without 
Gaussian correction as well as the fit of the Gamma distribution 
to the high-probability region ($P(S)>10^{-3}$, 
$S=[R_\mathrm{min}=-40:R_\mathrm{max}=23]$) only.
        Inset: High probability-region.
        \label{fig:distributionCompareGlobalPang}}
\end{figure}

We are not aware of any previous large-deviation study for multiple or for pairwise global alignment.
Here we first show our results for pairwise global alignment.
For a direct comparison to the results of Pang et al. \cite{pang_statistical_2005}, simulations were performed for alignments with identical gap costs of ($\alpha=7$,$\beta=1$), see Fig.\ \ref{fig:distributionCompareGlobalPang}.
First we performed a fit of the Gamma distribution \eqref{eq:gammaDistribution}, which was found in Ref.\ \cite{pang_statistical_2005} to fit the data well, to the high-probability region $P(S)>10^{-3}$ of the obtained distributions.
For pairwise alignment of sequences of lengths $L=50,100,200$ we obtained parameters similar to the  sampling results \cite{pang_statistical_2005}, as shown in tab. \ref{tab:pang_fitparams}.
As visible in Fig.\ \ref{fig:distributionCompareGlobalPang}, this fit does not match the data in the tail of the distribution.
Extending the fit to regions with lower probabilities yielded significantly different parameter values.
Nevertheless, this leads to a strong deviation in the high-probability region.
For a more elaborate study one could also obtain data for the low-scoring side of the distributions (by means of negative ``temperatures'' in the large deviation simulations).
Nevertheless, the results indicate that the assumption of scores distributed according to the Gamma distribution is somewhat flawed.
Thus, we also applied a Gaussian correction
\begin{equation}
    p_{\mathrm{gc}}(S)=
    \begin{cases}
        \frac{\lambda^{\gamma}(S-\mu)^{\gamma-1} e^{-\lambda(S-\mu)}}
        {\Gamma(\gamma)}e^{\lambda_2 S^2}
        & S>\mu\\
      0 & S\leq \mu,
    \end{cases}
    \label{eq:gammaDistributionCorrected}
\end{equation}
with the Gamma distribution \eqref{eq:gammaDistribution} and the Gaussian correction with the parameter $\lambda_2$ indicating its strength.
Fig.\ \ref{fig:distributionCompareGlobalPang} shows the fits of the different distributions.
Clearly, the Gamma distribution with the Gaussian correction covers the distribution best over the whole probability range.

\begin{table}
    \centering
    \begin{tabular}{c c c c c}
        \hline\hline
        L & & Pang\cite{pang_statistical_2005} & high $P(S)$ \\\hline
        50 & $\gamma$ & 41.00 & 33(6) & \\
           & $\lambda$ & 0.63 & 0.60(5) & \\
           & $\mu$ & -84.60 & -66(5) & \\
           & $\frac{\chi^2}{\text{ndf}}$ & & 1.43 \\
        100 & $\gamma$ & 49.16 & 48(6) & \\
            & $\lambda$ & 0.55 & 0.57(4) & \\
            & $\mu$ & -115.44 & -95(5) & \\
          & $\frac{\chi^2}{\text{ndf}}$ & & 1.05 \\
        200 & $\gamma$ & 52.24 & 59(12) & \\
            & $\lambda$ & 0.44 & 0.50(5) & \\
            & $\mu$ & -153.25 & -123(12) & \\
          & $\frac{\chi^2}{\text{ndf}}$ & & 0.74 & \\
        \hline\hline

    \end{tabular}
    \caption{Parameters for the fit of the Gamma distribution to the obtained data for global multiple sequence alignments as found in \cite{pang_statistical_2005} and as found with the large-deviation simulations with the fit restricted to $P(S)>10^{-3}$.
\label{tab:pang_fitparams}}
\end{table}

\subsubsection{Multiple global sequence alignments}

\begin{figure}
    \includegraphics[width=0.9\columnwidth]{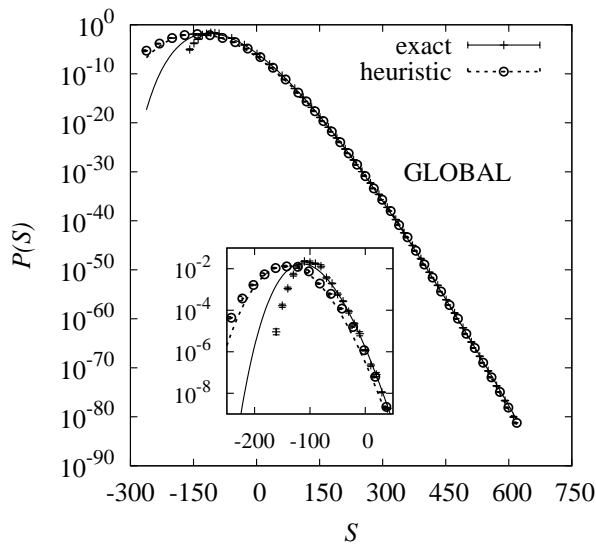}%
    \caption{
        Distributions obtained for the global alignment of $\Nseq=3$ 
sequences of length $L=40$ with the exact algorithm and the heuristics.
The lines show the fit of the \emph{uncorrected} Gamma 
function against the data.
        The inset shows the high-probability region, in which 
        both distributions differ significantly from each other and the 
fit for exact alignments deviates significantly from the data.
        \label{fig:compareHeuristicExact}}
\end{figure}

The results for the multiple global alignment of $\Nseq=3$ sequences 
of length $L=40$ are shown in Fig.~\ref{fig:compareHeuristicExact}.
Included are the distributions obtained with the progressive heuristics as well as with the computationally much more expensive exact algorithm.
The significant difference in the low-scoring region of the distributions is clearly visible.
In the high-scoring region the heuristics seems to work better, i.e., approaches the exact optimum alignments, which is visible in the better consensus between the two probability distributions.
Shown are also fits to the (uncorrected) Gamma distribution.
    The distribution fits only well for the distribution obtained with the heuristics, but not for the distribution obtained with the exact algorithm, i.e.~the Gaussian correction does not seem to be necessary for multiple alignments of $\Nseq>2$ sequences \emph{when using the heuristics}.
For $\Nseq=2$ the correction was found to be necessary in any case.
But for pairwise alignment, there is only one possible pair of sequences to be aligned, which is done by the dynamic programming algorithm.
The true optimal alignment is found.
In contrast, when using the heuristics, for $\Nseq>2$ sequences the first sequence pair is aligned with the dynamic programming algorithm.
The following sequences are aligned to this initial alignment.
If the sequences are very dissimilar, the initial alignment does not reflect the structure of the multiple alignment very well.
This means, especially in the low-scoring region scores lower than the true optimal scores are found with progressive alignment.
The different behaviour in the high-probability region between pairwise and multiple alignments is one possible explanation for the difference in parameter behaviour.
Generally, the exact algorithm is computationally too expensive for multiple sequence alignment and the heuristics is used.
It should be noted that for the score statistics this use of the heuristics renders the Gaussian correction unnecessary.
Further observations for global multiple sequence alignment were made for the case of using the heuristics if not stated otherwise.

\subsubsection{Comparison of pairwise and multiple sequence alignment}

\begin{figure}
    \includegraphics[width=0.9\columnwidth]{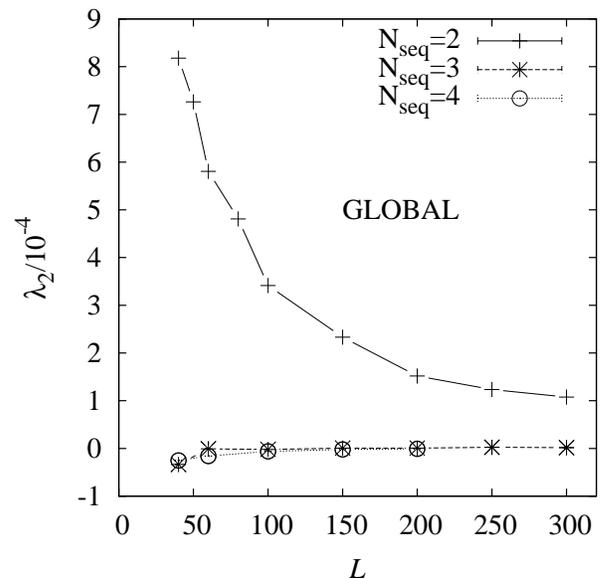}%
    \caption{
        The parameter $\lambda_2$ as found for the distributions of 
global alignment for different sequence lengths $L$ and different 
number of sequences $\Nseq$.
        Drawn lines are guides for the eye.
        \label{fig:l2_length_global}}
\end{figure}

Figure \ref{fig:l2_length_global} shows the values of fit parameter $\lambda_2$ for different sequence lengths and different number of sequences.
The most obvious observation to be made is that for $\Nseq=2$ the $\lambda_2$ values are positive, i.e., correcting the curvature of the distribution to lower probabilities, converging strongly towards zero.
For multiple alignments of $\Nseq>2$, however, $\lambda_2$ values are relatively close to 0.
This is compatible with the fact that a Gamma distribution without Gaussian correction seems to fit the data well when using the heuristics.
    Thus, when obtaining score distributions for significance analysis of sequence alignments, one actually needs to simulate multiple alignments 
explicitly, including large-deviations techniques.
It is not possible to deduce the distributions from the (large-deviation) results of pairwise alignment.

\begin{figure}
    \includegraphics[width=0.9\columnwidth]{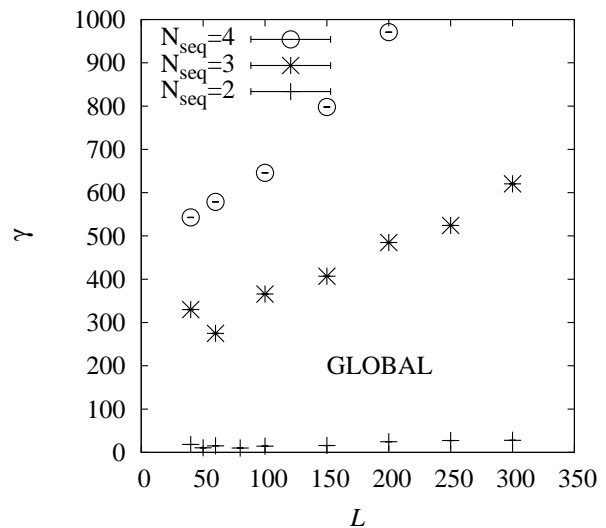}%
    \caption{
        The parameter $\gamma$ as found for the distributions for 
global alignment of different sequence lengths $L$ and different number 
of sequences $\Nseq$.
        \label{fig:gamma_length}}
\end{figure}

The behavior of the parameter $\gamma$ as function of sequence length is shown in Fig.~\ref{fig:gamma_length} for the three cases $\Nseq=2,3$ and $4$.
While the correction parameter $\lambda_2$ decreases with sequence length or is basically zero anyway, the parameter $\gamma$ increases.
For $\gamma \rightarrow \infty$ the Gamma distributions converges to the normal distribution.
Thus, for long enough sequences, the distribution of global alignment scores converges to a normal distribution.
This can be understood in the following way:
Neglecting gaps, which is in particular true for very large scores, the pairwise scores can approximately assumed to be i.i.d.~random numbers.
The overall sequence score is a sum over these values.
For $L \rightarrow \infty$ the central limit theorem holds, resulting in a Gaussian distribution.

\begin{figure}
    \includegraphics[width=0.9\columnwidth]{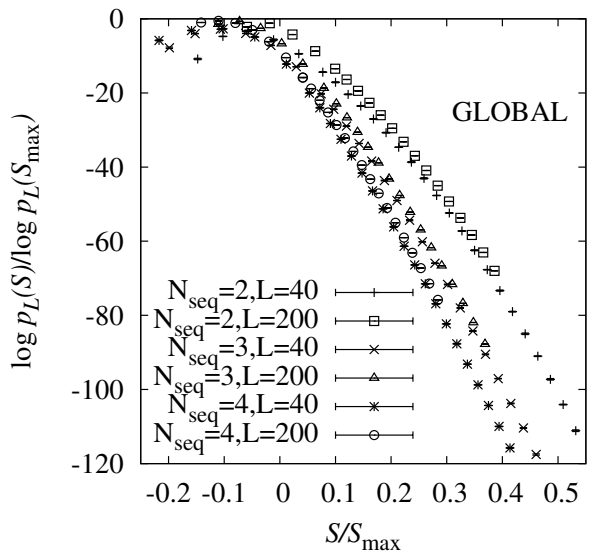}%
\caption{
    Distributions rescaled with \eqref{eq:significanceRelation} for 
different number of sequences $\Nseq$ and different sequence lengths $L$.
    The rescaled distributions coincide for low-probabilities and 
identical $\Nseq$ but differing $L$.
    \label{fig:compareN2-N4}}
\end{figure}

Same as for local sequence alignments, the probability distributions can be rescaled according to \eqref{eq:significanceRelation}.
The rescaled distributions are found in Fig.\ \ref{fig:compareN2-N4}.
Again, for identical number of sequences, the distributions coincide again for different sequence lengths $L$ in the low-probability region.
Again, there is a significant difference between the distributions for different number of sequences $\Nseq$, indicating that indeed significance analysis for multiple sequence alignments cannot be based on the results for pairwise alignments, i.e., dedicated multiple sequence alignment studies have to be performed explicitly, as done is this work.
Note that the initial deviation for higher probabilities seems more distinct than for local alignments, but it should be kept in mind that the sequence lengths shown for global alignments differ more than for local alignments, because more data was available.

\begin{figure}
    \includegraphics[width=0.9\columnwidth]{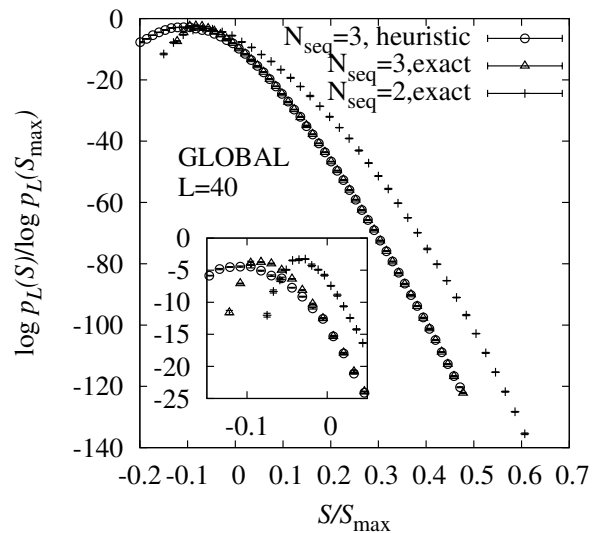}%
    \caption{
        Distributions obtained for global alignment of $\Nseq=2$ and $\Nseq=3$ sequences of length $L=40$ (with the exact algorithm as well as with the 
heuristics for $\Nseq=3$).
        The $\log(P)$ values are rescaled with \eqref{eq:significanceRelation}
 and the scores with $S/S_\text{max}$. 
        The inherit difference between $\Nseq=2$ and $\Nseq=3$ is obviously 
not based the use of an heuristics. The inset show the high-probability 
region, in which all three obtained distributions differ. In case of 
$\Nseq=3$ this difference is small compared to the exact algorithm
and is due to the bad performance of the heuristics 
in the low-scoring region.
        \label{fig:compareN2N3Rescaled}}
\end{figure}

The strong difference of the resulting distribution of pairwise and multiple alignments could be influenced by the use of a heuristics for multiple alignments.
For small sequence length, we were able to compare the results of both algorithms.
Figure \ref{fig:compareN2N3Rescaled} shows the rescaled distribution obtained for $L=40$ for pairwise alignment and in case of multiple sequence alignment with $\Nseq=3$ the rescaled distributions obtained with the heuristics as well as with the exact algorithm.
The distributions obtained from heuristics and exact algorithm, respectively, differ only in a small region, while the results from pairwise and multiple alignment differ everywhere.
Thus, the different development of distributions does not seem to be a result of the application of a heuristics.

\subsection{Comparison of local and global alignments}
\begin{figure}
    \includegraphics[width=0.9\columnwidth]{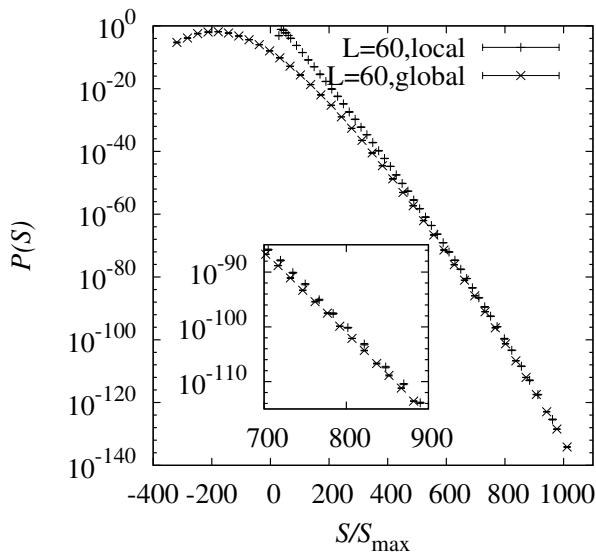}%
    \caption{
        Obtained distributions for local as well as global alignment of $\Nseq=3$ sequences of length $L=60$.
        The inset shows the high-scoring region, where both distributions approach each other.
        \label{fig:compareLocalGlobal}}
\end{figure}

Finally, the score distributions of local and global sequence alignment can be compared.
For mostly dissimilar sequences the selection of best-scoring subsequences can increase the alignment score.
Since the probability of the sampling of a specific sequence set is independent of the alignment algorithm, dissimilar sequence sets would score higher in local sequence alignment, where no negative scores $S$ are possible.
This means that the low-scoring region of the distribution for local alignments should be skewed towards higher alignment scores compared to the distribution of global alignments.
This can be observed in a comparison of score distributions with the different alignment types for the same sequence set properties, as shown in Fig.~\ref{fig:compareLocalGlobal}.
Here, as an example, the obtained distributions for $\Nseq=3$ sequences of length $L=60$ are shown.
As can be seen, the maximum of the distribution for global alignments is found for negative scores, the maximum for local alignments is positive (as local alignments can only have scores $S>0$).

On the other hand, most high-scoring global alignments, which have only few negative scoring residue-pairings or gaps, would not yield better scores by selecting (smaller) subsequences.
This means that in the high-scoring region, many sequence sets yield the same score in local and global alignment.
Therefore the distributions of local and global sequence alignment can be expected to agree for higher scores.
This can exactly be seen in Fig.~\ref{fig:compareLocalGlobal}.
This agreement might allow to estimate statistics of computationally expensive local alignments in the high-scoring region by means of the cheaper global alignment.
However, for practical applications in Molecular Biology, 
local alignment results typically in scores which are located in the intermediate region, where it differs from global alignments.
Thus, knowing the distribution of random-sequence scores for global alignment alone would not be sufficient to estimate the local-alignment score distribution here.

\section{Summary and Outlook}

We have studied the distribution of scores of multiple sequence alignments in the ensemble of i.i.d.~random protein sequences.
The statistical mechanics-based large deviation simulations used here allowed us to numerically measure the distribution of alignment scores in the biologically most relevant region of small probabilities, e.g., $\sim 10^{-100}$.
Data for these low-probability regions was not available before.

Analysis of the distribution for multiple \emph{local} sequence alignment showed that the Gumbel distribution alone does not describe the data as found previously for pairwise alignments \cite{wolfsheimer_local_2007}.
Further analysis showed that the suggested Gaussian correction improves the behaviour, but still is not sufficient to describe the whole distribution.
Nevertheless the results indicate that for a known distribution for a certain number of sequences $\Nseq$ and fixed sequence lengths $L$, distributions for other sequence lengths but the same $\Nseq$ can be found by rescaling. 

Also for pairwise \emph{global} sequence alignment, were previously only simple-sampling results were available, we could obtain the distribution of scores over a large range of the support.
We could reproduce the previous simple-sampling results \cite{pang_statistical_2005} only by restricting the fits to the high-probability region.
Extending the analysis to the low-probability region showed significant deviations of the Gamma distribution fit from the data.
Again, a Gaussian correction improves the fit of the distribution.
The comparison between multiple and pairwise global alignments revealed different characteristics of the fit parameters.
While the parameter $\lambda_2$ indicating the strength of the Gaussian correction was found to be positive but decreasing with sequence length for pairwise alignments, it was almost negligible (actually slightly negative) for multiple alignments, also converging towards zero.
The application of fast heuristics seems to be the reason for 
this difference in behaviour, because the behavior for multiple alignment 
was comparable to the behavior of pairwise alignment when using the progressive heuristics.
However, in the biologically relevant low-probability region the heuristics performs well and for small sequence length the distributions found with the exact algorithm and the heuristics do not differ.

It is interesting to note the behavior in the limit of infinitely
 long sequences, which is for biological applications not so relevant but of fundamental interest, in particular with respect to the analytical results.
Similar to previous studies for pairwise alignments, in this limit the distribution of scores are compatible with standard distributions:
For long sequences the score distribution of local alignments appears to converge to the Gumbel distribution, the score distribution of global alignments to a normal distribution.
Since in many actual applications, multiple alignments with more than ten sequences are not uncommon, simulation of sequence sets with a higher number of sequences 
($\Nseq>4$) appears sensible in future studies. 
Nevertheless, due to the computational complexity of the exact algorithm which depends strongly on $\Nseq$, this requires the design and implementation of faster algorithms, e.g., in particular efficient heuristics for multiple local sequence alignment.

Also it would be desirable, to extend these alignment simulations to more refined ensembles of random sequences, e.g., for biologically more specific ensembles like transmembrane proteins \cite{alignHMM2011}, or sequences with correlations.

\acknowledgments
PF acknowledges financial support from the German Science Foundation (DPG) within the Graduiertenkolleg GRK 1885.
The simulations were performed at the HPC Cluster HERO, located at the University of Oldenburg (Germany) and funded by the DFG through its Major Research Instrumentation Programme (INST 184/108-1 FUGG) and the Ministry of Science and Culture (MWK) of the Lower Saxony State. 

\bibliography{msa_pre}

\end{document}